\documentclass[times,10pt]{article} 
\usepackage{multicol}
\usepackage{graphicx} 
\usepackage{fullpage}

\begin{document}

\title{Locating Data in (Small-World?) Peer-to-Peer Scientific Collaborations}

\author{Adriana Iamnitchi$^1$ \and Matei Ripeanu$^1$ \and Ian Foster$^{1,2}$}

\date{$^1$Department of Computer Science, The University of Chicago \\
1100 E. 58th Street, Chicago, IL 60637, USA \\ 
$^2$Mathematics and Computer Science Division, Argonne National Laboratory\\ 
Argonne, IL 60439, USA \\ \vspace{0.3in} March 2002}

\maketitle

\begin{abstract}
Data-sharing scientific collaborations have particular
characteristics, potentially different from the current peer-to-peer
environments. In this paper we advocate the benefits of exploiting
emergent patterns in self-configuring networks specialized for
scientific data-sharing collaborations. We speculate that a
peer-to-peer scientific collaboration network will exhibit small-world
topology, as do a large number of social networks for which the same
pattern has been documented. We propose a solution for locating data
in decentralized, scientific,  data-sharing environments that exploits
the small-worlds topology. The research challenge we raise is: what
protocols should be used to allow a self-configuring peer-to-peer
network to form small worlds similar to the way in which the humans
that use the network do in their social interactions? 

\end{abstract}

\section{Introduction}

Locating files based on their names is an essential mechanism for
large-scale data sharing collaborations. A peer-to-peer (P2P) approach
is preferable in many cases due to its ability to operate robustly in
dynamic environments. 

Existing P2P location mechanisms focus on specific data sharing
environments and, therefore, on specific requirements: in Gnutella \cite{clip2},
the emphasis is on easy sharing and fast file retrieval, with no
guarantees that files will always be located. In Freenet \cite{clarke00}, the
emphasis is on ensuring anonymity. In contrast, systems such as CAN
\cite{ratnasamy01-2}, Chord \cite{stoica01} and Tapestry \cite{zhao01}
guarantee that files are always located, while 
accepting increased overhead for file insertion and removal.

Data usage in scientific communities is different than in, for
example, music sharing environments: data usage often leads to
creation of new files, inserting a new dimension of dynamism into an
already dynamic system. Anonymity is not typically a requirement,
being generally undesirable for security and monitoring reasons. 

Among the scientific domains that have expressed interest in building
data-sharing communities are physics (e.g., GriPhyN project \cite{griphyn}),
astronomy (Sloan Digital Sky Survey project \cite{sdss}) and genomics
\cite{hgenome}. The 
Large Hadron Collider (LHC) experiment at CERN is a proof of the
physicists' interest and pressing need for large-scale data-sharing
solutions. Starting 2005, the LHC will produce Petabytes of raw data a
year that needs to be pre-processed, stored, and analyzed by teams
comprising 1000s of physicists around the world. In this process, even
more derived data will be produced. 100s of millions of files will
need to be managed, and storage at 100s of institutions will be
involved.

In this paper we advocate the benefits of exploiting emergent patterns
in self-configuring networks specialized for scientific data-sharing
collaborations. We speculate that a P2P scientific collaboration
network will exhibit small-world topology, as do a large number of
social networks for which the same pattern has been documented.

We sustain our intuition by observing the characteristics of
scientific data-sharing collaborations and studying the sharing
patterns of a high-energy physics community (Section \ref{sec:sw-in-science}). In Section \ref{sec:locating-files-in-sw}
we propose a solution for locating data in decentralized, scientific,
data-sharing environments that exploits the small-worlds topology. The
research challenge we raise is: what protocols should be used to allow
a self-configuring P2P network to form small worlds similar to the way
in which the humans that use the network do in their social
interactions? While we do not have a complete solution, we discuss
this problem in Section \ref{sec:creating-a-sw}.

\section{Small Worlds in Scientific Communities}
\label{sec:sw-in-science}
In many network-based applications, topology determines
performance. This observation captivated researchers who started to
study large real networks and found fascinating results: recurring
patterns emerge in real networks \cite{albert02}. For example, social networks, in
which nodes are people and edges are relationships; the world wide
web, in which nodes are pages and edges are hyperlinks; and neural
networks, in which nodes are neurons and edges are synapses or gap
junctions, are all small-world networks \cite{watts99-book}. Two characteristics
distinguish small-world networks: first, a small average path length,
typical of random graphs (here 'path' means shortest node-to-node
path); second, a large clustering coefficient that is independent of
network size. The clustering coefficient captures how many of a node's
neighbors are connected to each other. One can picture a small world
as a graph constructed by loosely connecting a set of almost complete
subgraphs.  

The small world example of most interest to us is the scientific
collaboration graph, where the nodes are scientists and two scientists
are connected if they have written an article together. Multiple
studies have shown that such graphs have a small-world character in
scientific collaborations spanning a variety of different domains,
including physics, biomedical research, neuroscience, mathematics, and
computer science. 

Typical uses of shared data in scientific collaborations have
particular characteristics:
\begin{itemize}
\item \textit{Group locality}. Users tend to work in groups: a group of users,
although not always located in geographical proximity, tends to use
the same set of resources (files). For example, members of a science
group access newly produced data to perform analyses or
simulations. This work may result into new data that will be of
interest to all scientists in the group, e.g., for comparison. File
location mechanisms such as those proposed in CAN, Chord, or Tapestry
\cite{zhao01} do not attempt to exploit this behavior: each member of the group
will hence pay the cost of locating a file of common interest.

\item \textit{Time locality}. The same user may request the same file multiple times
within short time intervals. This situation is different, for example,
from Gnutella usage patterns, where a user seldom downloads a file
again if it downloaded it in the past. (We mention that this
characteristic is influenced by the perceived costs of storing
vs. downloading, which may change in time.)

\end{itemize}

It is the intuition provided by the small-world phenomenon in real
networks and the typical use of scientific data presented above that
lead us to the following questions. Let us consider the following
network: a node is formed of data and its provider (the scientist who
produced the data), and two nodes are connected if the humans in those
nodes are interested in each other's data. The first question is: is
this a small-world network? Based on the analysis of data sharing
patterns in a physics collaboration (presented in Section \ref{sec:data-sharing-in-physics}) we
speculate that this network will be a small world. Second, how can
such small-world topology be exploited for performance in the
data-sharing environments of interest to us? Finally, how do we
translate the dynamics of scientific collaborations into
self-configuring network protocols (such as joining the network,
finding the right group of interests, adapting to changes in user's
interests, etc.)?

We believe this last question is relevant and challenging in the
context of self-configuring P2P networks. We support this idea by
answering the second question: in Section \ref{sec:locating-files-in-sw} we sketch a file location
strategy that exploits the small-world topology in the context of
scientific data-sharing collaborations. Once we show that a
small-world topology can be effectively exploited, designing
self-configuring topology protocols to induce specific topology
patterns becomes more interesting.

\subsection{Data Sharing in a Physics Collaboration}

\label{sec:data-sharing-in-physics}

The D0 collaboration \cite{D0} involves hundreds of physicists from 18
countries that share large amounts of data. Data is accessed from
remote locations through a software layer (SAM
\cite{loebel-carpenter01}) that provides 
file-based data management. We analyzed data access traces logged by
this system during January 2002.

\begin{figure}[hpt]
\begin{center}
\includegraphics[width=4in]{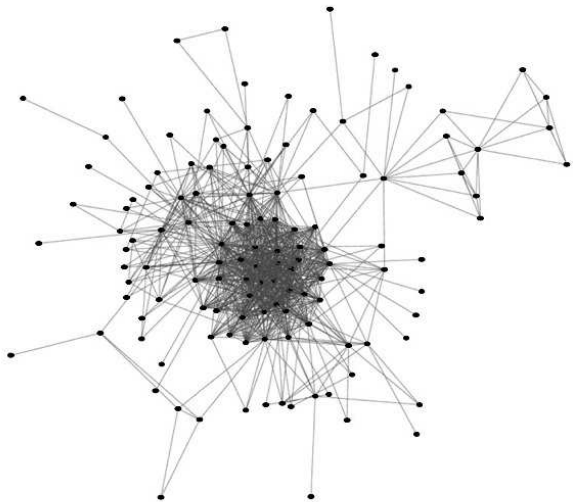}
\caption{The file-sharing graph of January 2002.}
\label{fig:sw-picture} 
\end{center}
\end{figure}

We considered the graph whose nodes are users and whose links connect
users that shared at least one file during a specified interval. We
found that the graphs generated for various interval lengths exhibit
small-world characteristics: short average path lengths and large
clustering coefficients. Although these graphs are relatively small
compared to our envisioned target (e.g., 155 users accessed files
through SAM in January), we expect similar usage patterns for larger
graphs. 

Table \ref{table} presents the characteristics of the graphs of users
who shared 
data within various time intervals ranging from 1 day to 30 days. The
small-world pattern is evident when comparing the clustering
coefficient and average path length with those of a random graph of
the same size (same number of nodes and edges): the clustering
coefficient of a small-world graph is significantly larger than that
of a similar random graph, while the average path length is about the
same.

  \begin{table}[hb]  
  \centering
  \caption{File-sharing graph characteristics for intervals from 1 to 30 days.}
  \bigskip 
  \begin{tabular}{|r|r|r|r|r|r|r|r|r|}
  \hline
  Interval & \multicolumn{2}{|c|}{Whole Graph} &
  \multicolumn{4}{|c|} {Largest Connected Component} &
  \multicolumn{2}{|c|}{Random Graph} \\
    \hline
  & \# Nodes & \# Links & \# Nodes & \# Links & Clustering & Path
  Lenght & Clustering & Path Lenght\\
\hline
      1 day & 20 &     38 &     12 & 34 &      0.827 &      1.61 &
      0.236 &     2.39 \\
      2 days &
     20 &
     77 &
     15 &
     75 &
     0.859 &
     1.29 &
     0.333 &
     1.68    \\
      7 days &
     63 &
     331 &
     58  &
     327 &
     0.816 &
     2.21 &
     0.097 &
     2.35     \\
      14 days &
     87 &
     561 &
     81 &
     546 &
     0.777 &
     2.56 &
     0.083 &
     2.30 \\
     30 days & 128 & 1046 & 126 & 1045 & 0.794 & 2.45 & 0.067 & 2.29 \\
 \hline
   \end{tabular}
 \label{table}

 \end{table}

\section{Locating Files in Small-World Networks}
\label{sec:locating-files-in-sw}

We consider an environment with potentially hundreds of thousands of
geographically distributed nodes that provide location information as
$<$logical filename, physical location$>$ pairs. 

Locating files in this environment is challenging because of scale and
dynamism: the number of nodes, logical files, requests, and concurrent
users (seen as file location requesters) may all be large. The system
has multiple sources of variation over time: files are created and
removed frequently; nodes join and leave the system without a
predictable pattern. In such a system with a large number of
components (nodes and files), even a low variation rate at the
individual level may aggregate into frequent group level changes.

We exploit the two environmental characteristics introduced in Section
2---group and time locality---to advance our performance objective of
minimizing file location latency. We also build on our assumption that
small-world structures eventually emerge in P2P scientific
collaborations. 

Consider a small world of $C$ clusters, each comprising, on average, $G$
nodes. A cluster is defined as a community with overlapping data
interests, independent of geographical or administrative
proximity. Clusters are linked together in a connected network. In
this structure, we combine information dissemination techniques with
request-forwarding search mechanisms: location information is
propagated aggressively within clusters, while inter-cluster search
uses request forwarding techniques. 

We chose gossip \cite{kermarrec00} as the information dissemination mechanism: nodes
gossip location information to other nodes within the
cluster. Eventually, with high probability, all nodes will learn about
all other nodes in the cluster. They will also know, with high
probability, all location information provided by all nodes within the
cluster. Hence, a request addressed to any node in the cluster can be
satisfied at that node, if the answer exists within the cluster. 

A request that cannot be answered by the local node is forwarded to
other cluster(s), by unicast, multicast, or flooding. Ideally,
clusters can organize themselves dynamically in search-optimized
structures, thus allowing a low cost inter-cluster file
retrieval. Since any node in a cluster has all information provided in
that cluster, the search space reduces from $C \times G$ to $C$.  

In this context, nodes need to store the total amount of information
provided by the cluster to which they belong. In order to reduce
storage costs, we use a compact, probabilistic representation of
information based on Bloom Filters (Section \ref{sec:bloom-filters}). Nodes can trade off
the amount of memory used for the accuracy in representing
information.

Each node needs to have sufficient topology knowledge to forward
requests outside the cluster. Not every node needs to be connected to
nodes from remote clusters, but, probabilistically, every node needs
to know a local node that has external connections. The question of
how to form and maintain inter-cluster connections pertains to the
open question we raise in this paper and discuss in Section
\ref{sec:creating-a-sw}: what 
topology protocols can induce the small-world phenomenon?

\section{Gossiping Bloom Filters for Information Dissemination}

In this section we briefly explain how we use the mechanisms mentioned
above: gossip for information dissemination and Bloom filters for
reducing the amount of communication. We also provide an intuitive
quantitative estimation of the system we consider.  

\subsection{Gossip Mechanism}

Gossip protocols have been employed as scalable and reliable
information dissemination mechanisms for group communication. Each
node in the group knows a partial, possibly inaccurate set of group
members. When a node has information to share, it sends it to a number
of $f$ nodes (fanout) in its set. A node that receives new information
will process it (for example, combine it with or update its own
information) and gossip it further to $f$ nodes chosen from its set. 

We use gossip protocols for two purposes: (1) to maintain accurate
membership information in a potentially dynamic cluster and (2) to
disseminate file location information to nodes in the local
cluster. We rely on soft-state mechanisms to remove stale information:
a node not heard about for some time is considered departed; a logical
file not advertised for some time is considered removed. 

\subsection{Bloom Filters}
\label{sec:bloom-filters}
Bloom filters \cite{bloom70} are compact data structures used for probabilistic
representation of a set in order to support membership queries ("Is
element $x$ in set $X$?"). The cost of this compact representation is a
small rate of false positives: the structure sometimes incorrectly
recognizes an element as member of the set.  

Bloom filters describe membership of a set A by using a bit vector of
length $m$ and $k$ hash functions, $h_1$, $h_2$, ..., $h_k$ with
$h_i:X \rightarrow {1..m}$. For a
fixed size ($n$) of the set to be represented, the tradeoff between
accuracy and space ($m$ bits) is controlled by the number of hash
functions used ($k$). The probability of a false positive is: 

$$ p_{err} \approx (1 - e^{-kn/m})^k $$

Here $p_{err}$ is minimized for $m/n \ln{2}$ hash functions. In practice, however, a
smaller number of hash functions is used: the computational overhead
of each additional hash function is constant while the incremental
benefit of adding a new hash function decreases after a certain
threshold. Experience shows that Bloom filters can be successfully
used to compress a set to 2 bytes per entry with false positive rates
of less than 0.1\% and lookup time of about 100$\mu$s.

A nice feature of Bloom filters is that they can be built
incrementally: as new elements are added to a set, the corresponding
positions are computed through the hash functions and bits are set in
the filter. Moreover, the filter expressing the reunion of multiple
sets is simply computed as the bit-wise OR applied over the
corresponding filters. 

Bloom filters can be compressed when transferred across the network
and, in this case, filter parameters can be chosen to maximize
compression rate, as shown in \cite{mitzenmacher01}.

\subsection{Advantages of Building the System around Shared Data
Interests}

We model this system built on group and time locality assumptions as
follows:

\begin{enumerate}
\item \textit{Zipf distribution for request popularity}. In Zipf distributions, the
number of requests for the $k$-th most popular item is proportional to
$k^{-\alpha}$, where $\alpha$ is a constant. Zipf distributions are
widely present in the 
Internet world. For example, the popularity of documents requested
from an Internet proxy cache (with $ 0.65<\alpha<0.85 $), Web server
document popularity ($0.75<\alpha<0.85$), and Gnutella query popularity
($ 0.63<\alpha<1.24 $) all exhibit Zipf distributions. For our problem we
assume that file popularity in each cluster (group) follows a Zipf
distribution. 

\item \textit{Locality of interests}. As discussed above, clusters are formed based
 on shared interest. We therefore assume that information on the most
 popular files is available within the cluster and only requests for
 not-so-popular files are forwarded. 

\end{enumerate}

\begin{figure}[hpt]
\begin{center}
\includegraphics[width=3.4in]{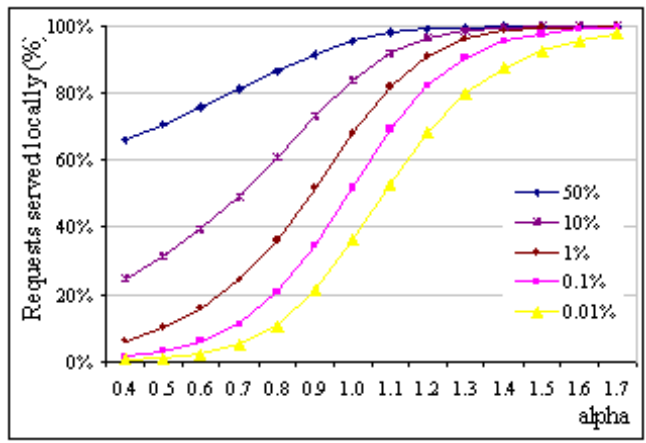}
\caption{Fraction of requests served locally (by one member of the
group) assuming various values of $\alpha$.}
\label{fig:zipf} 
\end{center}
\end{figure}

With these assumptions, we can estimate the fraction of file requests
served by the group as a function of the distribution parameter $\alpha$ and
the fraction of files about which the group maintains information. For
example, as Figure \ref{fig:zipf} shows, 68\% of all requests are served by the
group when information about only top 1\% most popular files is
available at group level, for $\alpha = 1$. Figure \ref{fig:zipf} strongly emphasizes the
need for efficient, interest-based cluster creation.

We estimate 100s of clusters with 1,000s of nodes in a cluster,
sharing information on about 10 million files per cluster. Using Bloom
filters, for 0.1\% false positives rate, each node needs 2 bytes per
file or 20MB of memory to store information about all files available
in the cluster. Assuming a 10-day average lifetime for a file at a
node, and a self-imposed threshold of 0.1\% false positives, then the
generated traffic needed to maintain this accuracy level within the
cluster can be estimated at about 24 KBps at each node.

False negatives may have two sources: the probabilistic information
dissemination mechanism and inaccuracy in the inter-cluster search
algorithm. By appropriately tuning the gossip periodicity and fanout,
the system can control the rate of false negatives by increasing
communication costs. 

\section{Creating a Small World}

\label{sec:creating-a-sw}

The question raised and not answered in this paper is: what protocols
should be used for allowing a self-configuring network to reflect the
small-world properties that exist at the social (as in a scientific
collaboration) level? There are at
least two ways to attempt to answer this question. The first approach
is to look at existing small worlds and to identify the
characteristics that foster the small-world phenomenon. The second
approach is to start from theoretical models that generate small
worlds \cite{watts99-book} and mirror them into protocol design.

The Gnutella network is an interesting case study as it is a P2P
self-configuring technological network that exhibits (moderate) small-world
characteristics \cite{jovanovic01}. How are the small-world characteristics
generated? One possible answer is that the social network formed by
the Gnutella users reflects its small-world patterns onto the
technological network. While this is not impossible, we observe that a
user has a very limited contribution to the Gnutella network
topology. Hence, we believe the social influence on the Gnutella
topology is insignificant. 

More significant for the small-world phenomenon may be Gnutella's
network exploration protocol based on \texttt{ping} and \texttt{pong} messages: a \texttt{ping}
is sent to all neighbors and each neighbor forwards it further to its
own neighbors, and so on. The \texttt{pong} messages return on the same path,
allowing a node to learn of its neighbor's neighbors, and hence to
improve clustering. However, the influence of this mechanism is
limited by the (comparatively) small number of connections per
node. This fact explains why, despite an aggressive exploration of the
network, the clustering coefficient in Gnutella is not large (e.g., it
is an order of magnitude lower than the clustering coefficients in
coauthorship networks). 

The theoretical model for building small-world graphs
\cite{watts99-book} starts from 
a highly clustered graph (e.g., a lattice) and randomly adds or
rewires edges to connect different clusters. This methodology would be
relevant to us if we had the clusters already formed and connected.
Allowing clusters to form dynamically based on shared interests,
allowing them to learn about each others, to adapt to users' changing
interests (e.g., divide or merge with other clusters) are parts of the
problem we formulate and do not answer. However, let us assume that
clusters form independently based on out of band information (the way
the Gnutella network forms) and let us assume further that they do
eventually learn about each other. Possible approaches for
transforming a loosely connected graph of clusters into a small world
(hence, with small average path length) are:
\begin{enumerate}
\item The hands-off approach: random graphs have small average path
length. It is thus intuitive that "randomly" connected clusters will
form a small world. 

\item The centralized approach at the cluster level: in each cluster, one or
multiple nodes are assigned the task of creating external
connections. 

\item The agent-based approach: allow an agent to explore the network and
rewire it where necessary. This approach is usually rejected due to
associated security issues. 

\end{enumerate}

\section{Summary}
We studied the file location problem in decentralized,
self-configuring P2P networks associated with scientific data sharing
collaborations. A qualitative analysis of the characteristics of these
collaborations, quantitative analysis of file sharing information from
one such collaboration, and previous analyses of various social
networks lead us to speculate that a P2P scientific collaboration may
benefit from a small-world topology. We sketch a mechanism for
low-latency file retrieval that benefits from the particularities of
the scientific collaboration environments and a small-world
topology. While we do not provide a solution for building topology
protocols flexible enough to resemble the dynamics and patterns of
social interactions, we stress the relevance of this problem and we
discuss some possible directions for research.  

\section*{Acknowledgements}
We are grateful to John Weigand, Gabriele Garzoglio, and their
colleagues at Fermi National Accelerator Laboratory for their generous
help. This work was supported by the National Science Foundation
under contract ITR-0086044.

\end{document}